\documentclass[11pt,a4paper]{article}
\usepackage{fullpage}
\usepackage[T1]{fontenc}
\usepackage[utf8]{inputenc}
\usepackage[colorlinks=true,allcolors=blue]{hyperref}
\usepackage{amsmath, amsthm, graphicx,color,subcaption, enumerate, bm, cleveref, enumitem, array, tcolorbox, mathtools, booktabs}
\usepackage{thm-restate}
\usepackage[table]{xcolor}
\usepackage{berasans, beramono}
\usepackage[charter]{mathdesign}
\usepackage{tabularx}
\usepackage{multicol, multirow}
\usepackage{ragged2e}
\usepackage{lineno}
\usepackage[mathcal]{eucal}
\usepackage{titlesec}
\usepackage{caption}
\titlespacing{\paragraph}{0pt}{1ex}{1em}

\newtcolorbox{algorithmbox}[1]{
  colback=white,
  colframe=black,
  colbacktitle=black!85,
  coltitle=white,
  fonttitle=\bfseries,
  title={#1},
  boxrule=0.8pt,
  arc=0pt,
  left=6pt,
  right=6pt,
  top=6pt,
  bottom=6pt
}

\makeatletter
\@ifundefined{newcounteralias}{}{%
  \renewcommand\thmt@autorefsetup{%
    \@xa\def\csname\thmt@envname autorefname\@xa\endcsname
      \@xa{\thmt@thmname}%
  }%
}
\makeatother

\newtheorem{theorem}{Theorem}

\newtheorem{lemma}[theorem]{Lemma}
\newtheorem{corollary}[theorem]{Corollary}
\newtheorem{definition}[theorem]{Definition}

\newtheorem{problem}{Problem}

\crefname{observation}{observation}{observations}
\Crefname{observation}{Observation}{Observations}
\crefname{problem}{problem}{problems}
\Crefname{problem}{Problem}{Problems}

\newcommand{\R}{\mathbb{R}}

\newcommand{\pre}{\mathrm{pre}}
\newcommand{\eps}{\varepsilon}

\newcommand{\OO}{\widetilde{O}}

\newcommand{\pVor}{\partial\mathrm{Vor}}
\DeclareMathOperator{\Vor}{Vor}
\DeclareMathOperator{\VD}{VD}

\DeclareMathOperator*{\argmax}{arg\,max}

\usepackage{authblk}
\usepackage{orcidlink}

\title{Faster Planar Diameter using Pseudoline Envelopes
\thanks{This research was supported by  the Austrian Science Fund (FWF) under grant DOI 10.55776/Z422.
For open access purposes, the authors have applied a CC BY public copyright license to any author-accepted manuscript version arising from this submission.
This research was partially funded by Israel Science Foundation grant No. 843/23. }}

\author[1]{Da Wei Zheng}
\affil[1]{Institute of Science and Technology Austria (ISTA), Austria\\
\href{mailto:dzheng@ista.ac.at}{dzheng@ista.ac.at}
\orcidlink{0000-0002-0844-9457}
}

\author[2]{Daniel Prigan}
\affil[2]{Reichman University, Israel\\
\href{mailto:daniel.prigan@post.runi.ac.il}{daniel.prigan@post.runi.ac.il}
\orcidlink{0009-0000-9285-3227}
}
\date{}

\begin{document}

\maketitle

\begin{abstract}
We give an improved algorithm for weighted directed planar diameter that runs in $\widetilde{O}(n^{3/2})$ time.
This improves upon the $\widetilde{O}(n^{5/3})$ time algorithm for weighted directed planar diameter of Gawrychowski, Kaplan, Mozes, Sharir, and Weimann [SIAM J. Comput. 2021] and the recent $\widetilde{O}(n^{8/5})$ time algorithm for unweighted undirected planar diameter of Fredslund-Hansen, Mozes, and Weimann [SOSA 2027].

The main new result is an improved data structure for computing the farthest vertex when given feasible planar distances to the vertices on the outer face.
The critical insight is the fact that normalized distances to a face behave like \emph{pseudolines}, for which we can leverage range searching and upper envelope data structures.
Using the farthest vertex data structure, we are able to compute all vertex eccentricities and diameter in $\OO(n^{3/2})$.
\end{abstract}

\section{Introduction}
Let $G$ be a directed graph with
$n$ vertices, real edge weights 
$w:E\to\mathbb{R}$, and no negative weight cycles.
Write $V(G)$ and $E(G)$ for the vertex and edge sets of $G$, and $d_G(u,v)$ for the length of a shortest directed $u$-to-$v$ path. 
We set $d_G(u,v)=+\infty$ when $v$ is unreachable from $u$, and $d_G(u,u)=0$.
In particular, $d_G(u,v)$ need not equal $d_G(v,u)$ in a directed graph. 
The outgoing eccentricity of $u$ and the diameter of $G$ are
\[
  \operatorname{ecc}^{+}_G(u)
    =\max_{v\in V(G)}d_G(u,v),
  \qquad
  \operatorname{diam}(G)
    =\max_{u\in V(G)}\operatorname{ecc}^{+}_G(u).
\]
Throughout the paper, eccentricity means outgoing eccentricity.

Computing the diameter of the graph is a fundamental problem in graph algorithms. For sparse graphs with $n$ vertices, $m=O(n)$ edges, and no negative weight edges, the simple algorithm of performing single source shortest path (SSSP) computation from every vertex runs in $\OO(n^2)$ total time\footnote{Throughout the paper we use $\OO(\cdot)$ notation to hide polylog factors.
} using Dijkstra's algorithm.
Roditty and Vassilevska Williams~\cite{RodittyW13} showed that assuming the strong exponential time hypothesis (SETH) and the orthogonal vectors (OV) hypothesis,
there is no $O(n^{2-\eps})$ time algorithm for any $\eps>0$.
Since obtaining a truly subquadratic time is unlikely for general graphs, significant attention has turned to computing exact diameter in more structured families of graphs, such as bounded treewidth graphs~\cite{CabelloK09,AbboudWW16, Husfeldt16, BringmannHM20}, planar graphs~\cite{Cabello19,GawrychowskiKMS21,FredslundHansenMW26}, minor-free graphs~\cite{DucoffeHV22,LeW24,KarczmarzZ25, ChanCGKLZ25, Zheng26}, and geometric intersection graphs~\cite{BringmannKKNP22, DurajKP24, Chang0024, ChanCGKLZ25, ChanCGKLZ26SoCG, ChanCGKLZ26ICALP}.

In a celebrated 2017 breakthrough, Cabello~\cite{Cabello19} showed that diameter in weighted directed planar graphs can be computed in $\OO(n^{11/6})$ running time. 
Gawrychowski, Kaplan, Mozes, Sharir and Weimann~\cite{GawrychowskiKMS21} subsequently improved the running time to $\OO(n^{5/3})$. 
Central to both results is the understanding of additively weighted Voronoi diagrams for planar distances.

Voronoi diagram techniques crucially exploited planarity and thus did not readily extend to other graph classes.
Progress in broader graph classes instead came from exploiting the combinatorial structure of distances through bounded VC-dimension.
Ducoffe, Habib, and Viennot~\cite{DucoffeHV22} first showed that subquadratic time diameter computation was possible in unweighted undirected minor-free graphs with a different structural technique involving distance VC-dimension.
In a different approach, using the distance profile set systems that were shown to have bounded VC-dimension in planar graphs by Li and Parter~\cite{LiP19}, 
Le and Wulff-Nilsen~\cite{LeW24} developed normalized distance profiles to obtain subquadratic diameter algorithms in unweighted minor-free digraphs.
Zheng~\cite{Zheng26} recently extended the VC-dimension
approach to handle minor-free digraphs with real edge weights.

Distance profile techniques have led to renewed progress on planar graphs.
Recently, Fredslund-Hansen, Mozes, and Weimann \cite{FredslundHansenMW26} improved the running time for unweighted undirected planar graphs to $\OO(n^{8/5})$ time by strengthening the Li and Parter~\cite{LiP19} bound for planar distance profiles.
This leaves a gap between $\OO(n^{5/3})$ time for weighted directed planar graphs and $\OO(n^{8/5})$ time in unweighted undirected planar graphs. Neither exponent seems to be optimal, raising the question of how far either bound can be improved.

Abboud, Mozes, and Weimann explicitly asked whether planar diameter admits
an $n^{1+o(1)}$-time algorithm or whether a superlinear lower
bound can be proved~\cite{AbboudMW23}.
At the 2016 Dagstuhl seminar \textit{Structure and Hardness in P}, S\o ren Dahlgaard suggested as an open problem that $n^{3/2-o(1)}$ conditional hardness was a possible target for planar diameter~\cite[Section~4.9]{LewensteinPW17}.
The exponent $3/2$ also has a natural algorithmic motivation:
the planar separator theorem of Lipton and
Tarjan~\cite{LiptonT79} guarantees a balanced separator with
$O(\sqrt n)$ vertices, 
and it is natural to need to consider all $O(n^{3/2})$ vertex--separator pairs in an algorithm to compute diameter.
This makes $n^{3/2}$ time a natural milestone for planar diameter upper and lower bounds.

\subsection{Results}
We reach this natural $n^{3/2}$ milestone for weighted planar diameter. 
\begin{theorem}\label{thm:diam-main}
There is an $\OO(n^{3/2})$ time algorithm for computing the diameter and all vertex eccentricities of a real-weighted directed $n$-vertex planar graph with no negative weight cycles.
\end{theorem}

The algorithm rests on a new structural theorem about planar
shortest-path distances: 
\emph{normalized distance profiles form a family of pseudolines}.
Fix a face of a plane graph $H$ with vertices $s_1, s_2, \dots, s_b$ ordered around the face.
For every vertex $x\in V(H)$, the normalized distance profile is the sequence
$\{d_H(s_i,x)-d_H(s_1,x)\}_{1\le i\le b}$.
We prove that any two such profiles change relative order at most
once as $i$ increases, meaning that this is a collection of pseudolines. See \Cref{fig:pseudolines} for an illustration. 
The most notable and surprising aspect of the new structural result is that $x$ is allowed to be anywhere in the graph.
The standard Monge property of planar graphs applies only to pairs of vertices on a single face.
This new structural result enables the usage of upper envelope and range searching
techniques for planar distances.
\begin{figure}[htb]
    \centering
    \includegraphics[page=2]{fig/Pseudolines.pdf}
    \caption{Left: an unweighted planar graph. Top right: distance profiles from four selected vertices, normalized by distance to $s_1$. Bottom right: Every pseudoline defined by distance profiles crosses every other pseudoline at most once.}
    \label{fig:pseudolines}
\end{figure}

A basic obstacle to applying range searching to graph distances
is the size of the separator.
The framework of Cabello and Knauer~\cite{CabelloK09} organizes
distances across a separator using orthogonal range searching,
with dimension depending on the number of separator vertices.
This dependence is effective for bounded-treewidth graphs that have constant sized balanced separators, 
but planar graphs can require balanced separators of size $\Theta(\sqrt n)$.
We show how to exploit planarity to avoid this. 
Instead we use Voronoi cells and the \emph{CCW-first} ordering of \cite{GawrychowskiKMS21} to organize the vertices into intervals, within each of which cell membership is determined by a single profile comparison.
The pseudoline property allows us to build efficient data structures for 
performing this operation. 

The resulting data structure removes a costly preprocessing step
in the planar Voronoi approach to diameter.
The construction of Gawrychowski et al.~\cite{GawrychowskiKMS21}
uses $\OO(Nb^2)$ preprocessing to compute the Voronoi diagram
of an $N$-vertex graph with $b$
sites. This data structure supports farthest-vertex queries in a Voronoi cell.
We present a new data structure that allows these farthest-vertex queries to be restricted to a prescribed subset $X\subseteq V(H)$, fixed during preprocessing:
for each site $s$, we seek the farthest vertex from $s$ among the
vertices of $X$ in its Voronoi cell.
We call $X$ the target set.
Writing $k=|X|$, our preprocessing takes only $\OO(N+bk)$ time.
This saves a factor of $b$ when all vertices are targets, while its
dependence on $k$ allows us to preserve distances in a large graph
and optimize over a much smaller region.
Combining this structure with the fast Voronoi construction of
Mozes and Prigan~\cite{MozesP2026} and a recursive planar
decomposition gives the $\OO(n^{3/2})$ running time.

\paragraph*{Geometry of distances between boundaries and interiors}
Monge inequalities have long been useful in computing distances in planar graphs, as in FR-Dijkstra~\cite{FakcharoenpholR06}. 
The connection between Monge matrices and pseudolines have been exploited
previously~\cite{Klawe92,KaplanMNS17,Chan21a} for matrix searching.
However, the standard Monge argument only applies to distance matrices of vertices that lie on a single face. 
Our new theorem can be viewed as a proof that the \emph{normalized} distance matrix between vertices on a single face and vertices anywhere else (after sorting rows and columns appropriately) is a \emph{totally monotone}, a condition that is weaker than Monge.

\subsection{Technical Overview}
We give an overview of our techniques to compute all vertex eccentricities, and hence diameter of a directed plane graph $G$. 
A key step is to do local computation on a strongly connected subgraph $H$ of $G$ with $N$ vertices and a set $B=\{s_1,\dots,s_b\}$ of $b$ vertices on a face that separates $H$ from some subset of vertices $Y\subseteq V(G)$.
To compute the farthest vertex in $H$ from a vertex $v\in Y$, we define a function $\omega:B\to\mathbb{R}$ where $\omega(s) = d_G(v, s)$ for $s\in B$, and find a vertex $x\in V(H)$ that maximizes $d_G(v, x)$. In notation this is:
\[
\max_{x\in V(H)}\min_{s\in B}
\{\omega(s)+d_H(s,x)\}.
\]
To do so, we utilize Voronoi diagrams.
\paragraph*{Voronoi Diagrams} 
For additive weights $\omega:B\to\mathbb{R}$, 
we say that the weights are \emph{admissible}, if for every $u,v\in B$, it holds that $\omega(u) + d_H(u,v) \ge \omega(v)$ so the triangle inequality is respected. Note that shortest path distances in any graph satisfy triangle inequality and are thus admissible.
Define the \emph{Voronoi cell} of a boundary vertex $s\in B$ as
\[ \Vor(s) := \left\{ v\in V(H): s=\operatorname*{arg\,min}_{s_i\in B} \{\omega(s_i)+d_H(s_i,v)\} \right\}
\]
For simplicity we assume that no ties occur
so these cells partition $V(H)$ and form the
\emph{additively weighted Voronoi diagram}
$\VD(H,B,\omega)$. A Voronoi cell is represented in the dual graph, and for any $s\in B$, the boundary of the Voronoi cell $\Vor(s)$ is a collection of dual edges and faces and is denoted by $\pVor(s)$. 
The diagram has a compact reduced dual representation of total complexity $O(b)$, and a represented edge may stand for a long path of underlying dual edges.
The complexity of the Voronoi cell $|\pVor(s)|$ is measured by the number of faces that contain a vertex of $\Vor(s)$ and vertices of at least two other Voronoi cells. 
A standard fact (see \cite{Cabello19} or \cite[Corollary~2.3]{GawrychowskiKMS21}) is that a Voronoi diagram of $b$ sites has total complexity $\sum_{s\in B} |\pVor(s)| = O(b)$.

\paragraph*{Farthest vertex in Voronoi cells} 
The Voronoi diagram gives us a way to compute the distances:
\[
\max_{x\in V(H)}\min_{s\in B}
\{\omega(s)+d_H(s,x)\}
=
\max_{s\in B}
\left\{
\omega(s)+
\max_{x\in\Vor(s)}d_H(s,x)
\right\}.
\]
Gawrychowski, Kaplan, Mozes, Sharir, and Weimann \cite[Theorem~1.1]{GawrychowskiKMS21} provide such a data structure for $N$-vertex planar graphs with no negative weight cycles and $b$ vertices lying on $O(1)$ faces. After $\OO(Nb^2)$ preprocessing time, their data structure constructs the diagram for any admissible additive weight $\omega$ in $\OO(b)$ time. 
It then computes the maximum distance from each active site \(s\) to a vertex in $\Vor(s)$ in $\OO(|\pVor(s)|)$ time. Thus, both constructing the diagram and computing all cell maxima take $\OO(b)$ total time.

The most expensive part is preprocessing, not query time. For each pair of boundary vertices, their construction preprocesses all possible bisectors as the difference between the two additive weights varies. A fixed pair of boundary vertices requires $\OO(N)$ time and space to handle.
Processing all $\Theta(b^2)$ pairs therefore incurs $\OO(Nb^2)$ preprocessing time \cite{GawrychowskiKMS21}. 
We avoid this pairwise preprocessing by using \emph{distance profiles} that we will define shortly. When the sites lie on a single face, our data structure supports searching for the maximum among a fixed target subset of vertices $X$ with preprocessing time sensitive to its size. The following is our main data structure contribution.

\begin{restatable}[Farthest Vertex Data Structure]{theorem}{mainds}\label{thm:main_ds}
Let $H$ be an $N$-vertex directed plane graph with real edge
lengths and no negative-length cycles. Fix a set 
$B=\{s_1, \dots, s_b\}\subseteq V(H)$ of $b=|B|$ vertices on a single face of $H$ and a subset $X\subseteq V(H)$ with $k=|X|$ vertices.
After $\OO(N+bk)$-time preprocessing, a query of admissible weights $\omega:B\to\mathbb{R}$ admits the computation of an implicit representation of $\VD(H,B,\omega)$ in $\OO(b)$ time.
This representation supports querying for $\max_{v\in\Vor(s)\cap X} d_H(s,v)$ for any $s\in B$ and a vertex (if any exist) witnessing the max in $\OO(|\pVor(s)|)$ time.
\end{restatable}
The target vertex set $X$ is fixed during preprocessing, while the additive weights $\omega$ are supplied with each query. In the diameter algorithm, the graph we are building the data structure on may be much larger than $|X|$, making this refined preprocessing time essential.

\paragraph*{Distance profiles and pseudolines}
Fix an arbitrary vertex, say  $s_1\in B$, where the vertices $B=\{s_1, s_2, \dots, s_b\}$ are ordered counterclockwise around a single face. Define the \emph{normalized distance profile} of a target $x\in X$ as 
\[ f_x^{s_1}(s_i):=d_H(s_i,x)-d_H(s_1,x),\qquad s_i\in B.\]
These were examined by Le and Wulff-Nilsen~\cite{LeW24} which were inspired by the patterns of Li and Parter~\cite{LiP19}. 
The crucial observation is that viewing the distance profile as a (discrete) curve going through the points $(1, f_x^{s_1}(s_1)),  (2, f_x^{s_1}(s_2)), \dots, (b, f_x^{s_1}(s_b))$, these curves satisfy a \emph{pseudoline property}: two curves change relative order at most once along the boundary. See \Cref{fig:pseudolines} for an illustration of this.
The proof of this follows from a variation of the result of Li and Parter \cite{LiP19}, who show that unweighted graphs avoid alternating patterns.
The pseudoline property indicates that these profiles behave similarly to abstract lines in the plane. In particular, this implies upper envelopes of $k$ pseudolines have complexity $O(k)$.

\paragraph*{Handling a query}
For each query, we construct the additively weighted Voronoi diagram with the recent fast construction algorithm of Mozes and Prigan~\cite{MozesP2026} in $\OO(b)$ time.
We then use the fact that the Voronoi edges of $\Vor(s_1)$ partition the \emph{CCW-first} ordering of the shortest-path tree $T_{s_1}$ rooted at $s_1$ into one circular interval per Voronoi edge, as observed in \cite{GawrychowskiKMS21}.
The cell $\Vor(s_1)$ induces a connected subtree of $T_{s_1}$, and each excursion of the tree traversal outside the cell visits an entire attached subtree before returning.
The outward edges of the cell occur in the order of its dual boundary, so all excursions through the same reduced Voronoi edge lie in its corresponding interval.
See \Cref{ssec:ccw-first} for details and \Cref{fig:voronoi-euler-intervals} for an illustration.
On the interval corresponding to an edge separating the Voronoi cells of $s_1$ and $s_i$, membership in the Voronoi cell of $s_1$ is determined by the single comparison $f_x^{s_1}(s_i)>\omega(s_1)-\omega(s_i)$.
For an $x$ outside $\Vor(s_1)$ in the interval, the shortest path from $s_1$ to $x$ first leaves $\Vor(s_1)$ at a vertex $v\in\Vor(s_i)$.
Replacing the $s_1$-to-$v$ prefix by a shortest path from $s_i$ to $v$ yields $f_x^{s_1}(s_i)<\omega(s_1)-\omega(s_i)$, so the comparison with $s_i$ alone excludes every vertex outside $\Vor(s_1)$ in the interval.

We find the farthest vertex of $\Vor(s_1)\cap X$ within a CCW-first interval $I$ by binary searching the sorted distances $d_H(s_1,x)$ for $x\in X$.
For a candidate distance $D$, we consider targets whose CCW-first positions lie in $I$ and whose distances from $s_1$ are at least $D$.
A data structure on CCW-first positions and distances decomposes the candidate set into polylogarithmically many canonical subsets, whose upper envelopes are stored during preprocessing.
If the largest envelope value at $s_i$ exceeds $\omega(s_1)-\omega(s_i)$, the vertex $x$ corresponding to a maximizing profile $f_x^{s_1}$ belongs to $\Vor(s_1)$ and satisfies $d_H(s_1,x)\ge D$; otherwise, no candidate belongs to $\Vor(s_1)$.
The pseudoline property of normalized distance profiles shows that these upper envelopes have linear complexity.
Standard range searching then achieves $\OO(bk)$ preprocessing for the $k$ target vertices, in addition to $\OO(N)$ shared global preprocessing.
The reduced Voronoi diagram has $O(b)$ Voronoi edges, each representing a bisector between two sites and defining one interval in the CCW-first ordering rooted at each of the two sites.
Thus, processing the resulting $O(b)$ intervals in $\OO(1)$ time per interval gives $\OO(b)$ total query time.

\paragraph*{From data structure queries to diameter.}
We apply the data structure within a recursive $\mathbf{r}$-division with few holes~\cite{KleinMS13}. 
The important insight is that it suffices to handle distances between vertices on successive levels of the $\mathbf{r}$-division. This way, we can separately consider the holes of each piece of the division in every level of the hierarchy as a single face to apply \Cref{thm:main_ds}.
Pairs of vertices between these successive levels 
and the remaining pairs at the leaves account for every pair of vertices in the graph. Each vertex maintains the greatest distance found, yielding eccentricities and hence the diameter.

\subsection{Organization}
We begin with preliminaries in \Cref{sec:preliminaries}, before presenting the full algorithm for diameter and eccentricities using our data structure of \Cref{thm:main_ds} in \Cref{sec:diameter-algorithm}. 
Next, we analyze the structural connection between distance profiles and pseudolines in \Cref{sec:pseudolines} before proving \Cref{thm:main_ds} in \Cref{sec:datastructure}. 
\section{Preliminaries}\label{sec:preliminaries}
Unless stated otherwise, $G$ is a directed planar graph with
$n=|V|$ vertices and $O(n)$ edges with a weight function $w(e) = w(u,v)$ for each edge $(u,v)$. We assume $G$ has no negative weight cycles.
For a subgraph $H\subseteq G$ with $N$ vertices, we use $V(H)$ to denote the vertices of $H$, and $E(H)$ to denote the edges of $H$. We designate a face of $H$ as the outer face and let $B$ be the set of vertices on the boundary.
A \emph{plane graph} is a planar graph with a fixed embedding.
Connectivity, faces, and planar duality refer to the underlying
undirected embedded graph.
We write 
$e^*$ and $f^*$ for the dual edge of a primal edge $e$ and the dual vertex of a face $f$.

\paragraph{Simplifying assumptions.}
For the weighted algorithm for computing eccentricities, we will assume that the undirected version of the input planar graph is simple and maximal (thus triangulated), and we have weighted directed edges going in both directions (so the graph is strongly connected). We justify these assumptions as follows.
We can remove self-loops and for each ordered pair of adjacent vertices, retain only a shortest directed edge.
Define $M=\sum_{e\in E(G)}|w(e)|.$
Every reachable pair has a shortest simple path of absolute value at most $M$.
Add edges to any embedding of the graph to make it maximal, 
giving length $2M+1$ to both directions of every added edge and to
each missing reverse directed edge.
Note that for any two vertices $u,v \in V(G)$ where $u$ can reach $v$, the shortest path distance did not change by the transformation. On the other hand if $u$ cannot reach $v$, the shortest path distance will be at least $2M+1-M=M+1$, as the sum of all original edges is at least $-M$. Thus if a shortest path from $u$ to $v$ is larger than $M$ in the augmented graph, then $u$ does not reach $v$ in the original graph.
Furthermore, we can assume that the lengths of all shortest paths are unique.
This can be done by standard symbolic perturbation of the edges.

\paragraph*{Recursive subdivisions}
A \emph{region} of a planar graph is an edge-induced embedded subgraph where we ignore edge orientations.
A \emph{division} covers the edges of the graph by regions, which may overlap
on their boundaries. A vertex shared by multiple regions is a
\emph{boundary vertex}. The boundary vertices of a region $R$ are denoted by
$\partial R$.
A \emph{hole} of $R$ is a face of $R$ that is not a face of the original
embedded graph.
An \emph{$r$-division} is a division with $O(n/r)$ regions, each with at most $r$
vertices and $O(\sqrt r)$ boundary vertices. It has \emph{few holes}
if each region has $O(1)$ holes and every edge shared by two regions
lies on a hole boundary in each of them~\cite{KleinMS13}.

A decomposition tree is a rooted hierarchy of regions whose root is
the entire graph and whose children cover their parent.
For a geometrically decreasing sequence
$\mathbf r=(r_m,\ldots,r_1)$, a \emph{recursive
$\mathbf r$-division with few holes} is such a hierarchy where each level is a $r_i$-division with few holes.

\begin{theorem}[Recursive divisions~{\cite[Theorems~2 and~3]{KleinMS13}}]
\label{thm:recursive-division}
For any geometrically decreasing sequence $\mathbf r=(r_m, \ldots, r_1)$ with $r_1$ larger than some absolute constant, a recursive $\mathbf r$-division with few holes can be computed in $O(n)$ time.
\end{theorem}

We use the connected regions produced by this construction
\cite[Section~2.5]{KleinMS13}.
At scale $r$, there are $O(n/r)$ regions, each of size $O(r)$,
with $O(\sqrt r)$ boundary vertices and $O(1)$ holes.

\paragraph*{Multiple-source shortest paths (MSSP).}
For a strongly connected planar graph $H$, let $T_s$ denote the shortest-path tree rooted at $s$, with tree edges directed away from $s$. 

We use the multiple-source shortest-path (MSSP) algorithm of Klein~\cite{Klein05}. For a fixed face $B$ of an $N$-vertex planar graph $H$, the algorithm moves the root of the
shortest-path tree around the boundary of $B$. Over the entire
execution, the shortest-path tree undergoes only $O(N)$ changes,
and a persistent representation gives access to the tree rooted
at each boundary vertex.
\begin{theorem}[MSSP~{\cite{Klein05}}, \cite{MozesW10}]
\label{thm:mssp}
Let $H$ be an $N$-vertex directed plane graph no negative weight cycles, and let $B$ be a distinguished face.
In $\OO(N)$ time and space, one can construct a data
structure that, given $s\in B$ and $v\in V(H)$, returns
$d_H(s,v)$ in $O(\log N)$ time.
\end{theorem}
Note that the original algorithm of \cite{Klein05} could only handle nonnegative edge weights. 
Mozes and Wulff-Nilsen~\cite{MozesW10} show that by vertex reweighting, this can be extended to any graph with no negative weight cycle with $O(N\log^2N/\log\log N)$ additional preprocessing time. 

\paragraph*{Distance oracle.}
We will need exact distances between arbitrary ordered pairs of vertices.
We use the distance oracle of Charalampopoulos et al.~
\cite{CharalampopoulosGLMPWW23}, with the improved construction of
Mozes and Prigan~\cite{MozesP2026}.

\begin{theorem}[Distance oracle~{\cite[Theorem~1.1]{MozesP2026}}]
\label{thm:distance_oracle}
Let $G$ be an $n$-vertex directed planar graph with real edge weights
and no negative weight cycles. In $n^{1+o(1)}$ time and space, one can
construct a data structure that, given any ordered pair
$(u,v)\in V(G)\times V(G)$, returns $d_G(u,v)$ in
$\log^{2+o(1)}n$ time.
\end{theorem}

\section{Weighted Eccentricities and Diameter Algorithm}\label{sec:diameter-algorithm}

We prove Theorem~\ref{thm:diam-main} by applying the Voronoi farthest vertex data structure of \Cref{thm:main_ds} separately within each hole of a recursive division.
The high level strategy is to construct a recursive $\mathbf{r}$-division with few holes and consider each parent-child pair of regions $(P,Q)$. 
For each hole $h$ of $Q$, we query distances from vertices of $Q$ to vertices of $P$ lying in $h$ or on its boundary. 

Recall that the input is an $n$-vertex planar graph $G$ that is strongly connected and triangulated.
We present the entire algorithm here for computing the eccentricities of all vertices.

\begin{enumerate}
    \item \textbf{Construct a global distance oracle.}
    Construct a single distance oracle on the entire graph $G$
    using \Cref{thm:distance_oracle}.
    This oracle is used in steps 4(b) and 4(c) of the algorithm.
    
    \item \textbf{Construct a recursive $\mathbf{r}$-division.}
    Construct a recursive $\mathbf{r}$-division with few holes of
    \Cref{thm:recursive-division}, with geometrically decreasing scales
    $\mathbf r = (r_m,\dots,r_1)$ and $r_1=L$ for a parameter $L$
    we will specify later.

    \item \textbf{Initialize the answers.}
    For every vertex $u\in V(G)$, let $A[u]$ denote its tentative
    eccentricity and $W[u]$ denote a vertex witnessing this value.
    Initialize $A[u]=0$ and $W[u]=u$.

    \item \textbf{Process distances.}
    For every parent-child pair $(P,Q)$ in the hierarchy, perform Steps~4(a) and~4(b) separately for each hole $h$ of $Q$, separating $Q$ from $P\setminus Q$, including the exterior hole when present.
    Let $B_h$ be the boundary walk around the hole $h$. (If the hole $h$ is not simple, e.g. because of pendant vertices or bridges, perform the standard transformation of adding an additional copy of each vertex for each occurrence in a facial walk and add zero weight edges to make the boundary simple.)
    Note that Step~4(c) is performed only when $Q$ is a leaf.
    \begin{enumerate}
        \item \textbf{Build the local data structures.}
        Define $H_{P,h}$ by retaining the part of $G$ drawn in $h$, together with its boundary $B_h$. Thus $V(H_{P,h})\cap V(Q)=V(B_h)$. (Note that $H_{P,h}$ does not contain $Q$).
        Using \Cref{thm:main_ds}, choosing $X=V(P)\cap V(H_{P,h})$, construct the implicit data structure for querying $\VD(H_{P,h},B_h,\omega)$ for $\omega$ defined by vertices within $Q$.
        
        \item \textbf{Handle child to parent distances.}
        For every vertex $u\in V(Q)$, query the global distance oracle of Step 1 to obtain the additive boundary weights $\omega_u(s)=d_G(u,s)$ for $s\in B_h$.
        Using the implicit data structure for
        $\VD(H_{P,h},B_h,\omega_u)$, query with the additive
        weights $\omega_u$ to compute the farthest vertex from
        $u$ in $\Vor(s)\cap V(P)$ for all $s\in B_h$,
        updating $A[u]$ and $W[u]$ if a farther vertex is found.
        The distances used within the Voronoi data structure
        remain $d_{H_{P,h}}$.

        \item \textbf{Handle distances within leaves.}
        If $Q$ is a leaf, query the global distance oracle for
        $d_G(u,v)$ for every ordered pair
        $(u,v)\in V(Q)\times V(Q)$.
        Update $A[u]$ and $W[u]$ if a farther vertex is found.
        These distances are computed in the full graph $G$ using the distance oracles of Step 1.
    \end{enumerate}
\end{enumerate}
At the end of the algorithm, we will have computed the eccentricity of every vertex $u\in V(G)$ as $A[u]$ and a witness to the eccentricity $W[u]$. The diameter can be computed as the maximum eccentricity.

\subsection{Correctness}
Fix a vertex $u\in V(G)$ with $u\in V(Q_u)$ for a leaf region $Q_u$. Suppose the eccentricity of $u$ is witnessed by a vertex $t\in V(G)$. We divide into two cases depending on whether $t\in Q_u$. In either case, we show that we have considered the distance from $u$ to $t$ as a candidate for eccentricity.
We also show that every candidate considered in Steps~4(b) and~4(c) is an exact distance in $G$, and hence cannot exceed the eccentricity of $u$; together with coverage, this proves correctness.

\paragraph*{Coverage of leaf distances} Suppose $t\in Q_u$. 
Step~4(c) queries the global distance oracle for every ordered pair of vertices in $V(Q_u)$. 
In particular, the oracle returns the exact distance $d_G(u,t)$, even if a shortest path leaves and reenters $Q_u$. 
This distance is considered as a candidate for the eccentricity of $u$.

\paragraph*{Coverage of non-leaf distances} Suppose that $t\not\in Q_u$. Then there must be some parent and child region $P$ and $Q$ with $u\in Q$,  and $t\in P\setminus Q$  where $h$ is the hole of $Q$ containing $t$, with boundary $B_h$.
The piece $Q$ can be found be found by walking up the hierarchy from $Q_u$ until we find that $t\in P$.
Step~4(b) queries the global distance oracle to obtain $\omega_u(s)=d_G(u,s)$ for every $s\in B_h$, and constructs an implicit representation of $\VD(H_{P,h},B_h,\omega_u)$.

The following argument applies to every processed parent-child pair $(P,Q)$, every hole $h$ of $Q$, and every source $u\in V(Q)$.
For every $v\in V(H_{P,h})$, splitting a shortest $u$-to-$v$ path at its last vertex of $B_h$ gives $d_G(u,v)=\min_{b\in B_h}\{\omega_u(b)+d_{H_{P,h}}(b,v)\}$, since the remaining suffix lies in $H_{P,h}$.
Thus, for every $v\in\Vor(s)$, the value $\omega_u(s)+d_{H_{P,h}}(s,v)$ equals $d_G(u,v)$, so we have considered correct shortest paths for all candidates considered.

Suppose $t\in \Vor(s)\cap V(P)$ for $s\in B_h$. 
The data structure of \Cref{thm:main_ds} returns the farthest vertex in $\Vor(s)\cap V(P)$, suppose it returned some vertex $v'$. 
Thus $v'$ satisfies the expression as follows.
\[ v' = \argmax_{v\in \Vor(s)\cap V(P)} \omega_u(s) + d_{H_{P,h}}(s, v) = \argmax_{v\in \Vor(s)\cap V(P)} d_{G}(u, s) +d_{H_{P,h}}(s, v) = \argmax_{v\in \Vor(s)\cap V(P)} d_G(u, v)\]
Thus we have considered a distance $d_G(u,v') \ge d_G(u, t)$ as a candidate for the eccentricity of $u$. 
Since $t$ witnesses the eccentricity of $u$, equality follows, and $v'$ is therefore globally farthest from $u$.

\subsection{Running time}
The global distance oracle of Step 1 is constructed once in $n^{1+o(1)}$ time using \Cref{thm:distance_oracle}.
Step 2 takes $O(n)$ time using \Cref{thm:recursive-division}. 

To analyze Step~4, consider a child region $Q$ at scale $r_i$ of the recursive $\mathbf{r}$-division, its parent $P$, and a hole $h$ of $Q$.
In particular, this means that both $P$ and $Q$ have size $O(r_i)$, and the boundary hole $B_h$ has size $b = O(\sqrt{r_i})$.
Step~4(a) applies \Cref{thm:main_ds} to $H_{P,h}$ of size $O(n)$, with $|B_h|=O(\sqrt{r_i})$ and $|X|=O(r_i)$.
This takes $\OO(n+r_i^{3/2})$ preprocessing time.

Step~4(b) makes $O(|V(Q)|\,|B_h|)=O(r_i^{3/2})$ oracle queries to obtain the additive weights, taking $O(r_i^{3/2}\log^{2+o(1)}n)=\OO(r_i^{3/2})$ time.
For each of the $O(r_i)$ vertices of $Q$, constructing $\VD(H_{P,h},B_h,\omega_u)$ takes $\OO(b)$ time by \Cref{thm:main_ds}.
Querying all Voronoi cells takes $\OO(\sum_{s\in B_h}|\pVor(s)|)=\OO(\sqrt{r_i})$ time
for each $\omega_u$, taking $\OO(r_i^{3/2})$ time in total.
Step~4(c) makes $O(|V(Q)|^2)=O(L^2)$ oracle queries per leaf, taking $\OO(L^2)$ time per leaf and $\OO(nL)$ time over all leaves, since all leaves have size at most $L$.
Using the fact that there are $O(n/r_i)$ many regions at scale $r_i$, and that each child region has $O(1)$ holes, so processing all its holes changes the bounds for Steps~4(a) and~4(b) by only a constant factor.
Therefore, the total running time is:

\begin{align*}
n^{1+o(1)}
+\OO\left(
    \frac{n}{L}\cdot L^2
    +\sum_{i=1}^{m}\frac{n}{r_i}(n+r_i^{3/2})
\right)
& =
n^{1+o(1)}
+\OO\left(
    nL+n^2\sum_{i=1}^{m}\frac{1}{r_i}
    +n\sum_{i=1}^{m}\sqrt{r_i}
\right)\\
& =
\OO\left(nL+\frac{n^2}{L}+n^{3/2}\right).
\end{align*}

The last equality uses that the scales are geometric, with minimum
$L$ and maximum $O(n)$. 
Setting $L=\sqrt{n}$ balances the terms $nL$ and $n^2/L$
and gives $\OO(n^{3/2})$ time.

\section{Pseudoline Families} \label{sec:pseudolines}
Here we give a careful definition of weak pseudoline families that allow for stretches of equality between the pseudolines, before showing that normalized distance profiles form a pseudoline family. Subsequently, we describe efficient data structures for constructing upper envelopes of distance profile pseudolines.
\subsection{Characterizing Pseudolines}
In this section we define what it means for a family of discrete functions to be pseudolines.
We say that the collection of functions $f_1,\dots, f_k:[b]\to \R$ is a \emph{weak pseudoline family}, 
if the ordered list of differences $\{f_i(x)-f_j(x)\}_{x\in [b]}$ switches signs at most once after omitting all zero entries. 
This induces a geometric \emph{belowness relationship}: we can say $f_i$ is \emph{weakly below $f_j$ at $x$}, or in notation $f_i(x) \prec f_j(x)$.
These functions are called pseudolines because if $f_i$ is initially below $f_j$, then there is at most one point where $f_i$ transitions to being \emph{weakly above} $f_j$, and $f_i$ will never be below $f_j$ afterwards.
We formally characterize this relation as it will be useful in our later algroithms.
\begin{definition} \label{def:order}
Let $b$ be a positive integer and consider $k$ discrete functions $f_1, \dots, f_k:[b] \to \R$. 
For $i,j\in [k]$ and $x\in [b]$\hspace{0.1em}, we can say that $f_i(x) \prec f_j(x)$ if the function $f_i$ lies \emph{weakly below} $f_j$ at $x$, meaning that one of the following holds:
\begin{enumerate}
    \item \textbf{Strictly below. } $f_i(x) < f_j(x)$.
    \item \textbf{Weakly below.} $f_i(x) = f_j(x)$ and  either:
    \begin{enumerate}
        \item $f_i(y) = f_j(y)$ for all $z< y\le x$ and $f_i(z) < f_j(z)$ for some $1\le z<x$.
        \item $f_i(y) = f_j(y)$ for all $1\le y < z$ and $f_i(z) < f_j(z)$ for some $x< z \le b$.
        \item $f_i(y) = f_j(y)$ for all $1\le y\le b$ and $i<j$.
    \end{enumerate}
\end{enumerate}
\end{definition}
This definition of a weak pseudoline family allows for stretches of equality and thus is slightly different from the standard definition of pseudolines in \cite{AgarwalS05}, but remark that they can be handled in essentially the same way (e.g. if you first make infinitesimal perturbations). We justify why we need this more cumbersome definition to handle these stretches of equality in an example in next section.

\subsection{Normalized Distance Profiles form a Pseudoline Family}
Recall that we consider a planar graph $H$, with a specific identified face $B = \{s_1, \dots,s_b\}$. Without loss of generality, we may assume that $B$ is the outer face. Recall also the definition of the \emph{normalized distance profiles}:
\[ f_x^{s_1}(s_i):=d_H(s_i,x)-d_H(s_1,x),\qquad s_i\in B.\]
These distance profiles have stretches of equality for technical reasons: there can be stretches of equality between two distance profiles that arise naturally from the shortest path topology. As an example consider two vertices $x,y\in V(H)$, where $y$ lies on the shortest path between $s_1$ and $x$ as well as on the shortest path between $s_2$ and $x$. In this case 
\begin{align*} f_x^{s_1}(s_2) 
&=d_H(s_2,x) - d_H(s_1,x) \\
&= d_H(s_2,y) +d_H(y, x) - \left(d_H(s_1,y) + d_H(y,x)\right)\\
&=d_H(s_2,y) - d_H(s_1,y) = f_y^{s_1}(s_2).  
\end{align*}

We prove here that normalized distance profiles $\{f_x^{s_1}\}_{x\in V(H)}$ form a weak pseudoline family by adapting the argument of Li and Parter \cite{LiP19}. Their argument was only for unweighted profiles, where they instead analyzed the discrete derivative of the distance profile $\{f_x^{s_1}(s_i)-f_x^{s_1}(s_{i-1})\}_{i=2,\dots,b}$. We generalize the argument to weighted profiles. 
\begin{lemma}\label{lem:pseudoline}
The set of normalized distance profiles $\{f_x^{s_1}\}_{x\in V(H)}$ forms a (weak) pseudoline family.
\end{lemma}
\begin{proof}
Suppose for sake of contradiction that there were two vertices $x,y\in V(H)$ where their normalized distance profiles $f_x^{s_1}$ and $f_y^{s_1}$ crossed twice. Interchanging $x$ and $y$ if necessary, there are
indices $1<i<j<k$ such that
\begin{equation}
\label{eq:normalized-alternation}
f_x^{s_1}(s_1)=f_y^{s_1}(s_1)=0, \qquad 
f_x^{s_1}(s_i)<f_y^{s_1}(s_i),\qquad
f_x^{s_1}(s_j)>f_y^{s_1}(s_j),\qquad
f_x^{s_1}(s_k)<f_y^{s_1}(s_k).
\end{equation}

\begin{figure}[!htbp]
    \centering
    \includegraphics[scale=0.94]{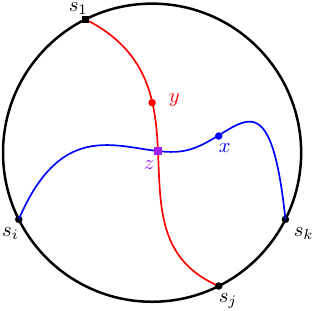}
    \caption{Example of a planar graph with $s_1, s_i,s_j,s_k$ in counterclockwise order around the boundary. There is no way for the undirected path from $s_1$ to $s_j$ through $y$ to avoid crossing the undirected path from $s_i$ to $s_k$ through $x$ without meeting at some vertex $z$. In notation of the proof of \Cref{lem:pseudoline} we would have $a=s_k$ and $b=s_1$.}
    \label{fig:liparter_pseudoline}
\end{figure}
Observe that the shortest path from $a$ to $x$, with
$a\in\{s_i,s_k\}$, must intersect some shortest path from $b$ to $y$, with
$b\in\{s_1,s_j\}$. This follows because $s_1, s_i, s_j, s_k$ appear in order on the boundary of $B$, and thus in the undirected graph, there cannot be an undirected path from $s_i$ to $x$ to $s_k$ that does not intersect the undirected path from $s_1$ to $y$ to $s_j$ since both lie on the same side of the outer face $B$ (see \Cref{fig:liparter_pseudoline}). 
Suppose that these two paths intersected at a vertex $z$.
\[
d_H(a,y)+d_H(b,x)\le
d_H(a,z)+d_H(z,y)+
d_H(b,z)+d_H(z,x)
= d_H(a,x)+d_H(b,y).
\]
Subtracting $d_H({s_1},x)+d_H({s_1},y)$ from both sides yields
\begin{equation}
f_y^{s_1}(a)+f_x^{s_1}(b)
\le f_x^{s_1}(a)+f_y^{s_1}(b).
\end{equation}
On the other hand, \eqref{eq:normalized-alternation} combined with the definition of $a$ and $b$ gives $f_x^{s_1}(a)<f_y^{s_1}(a)$ and $f_y^{s_1}(b)\le f_x^{s_1}(b)$. Adding these comparisons gives
\[
f_y^{s_1}(a)+f_x^{s_1}(b) >
f_x^{s_1}(a)+f_y^{s_1}(b)
\]
This is a contradiction, so we conclude that $\{f_x^{s_1}\}_{x\in V(H)}$ is a pseudoline family.
\end{proof}

\subsection{Data Structures for Upper Envelope of Distance Profile Pseudolines}
In this section we discuss basic computation of upper envelopes, and describe a data structure to quickly compute upper envelopes of distance profiles around different boundary vertices.

\paragraph*{Constructing upper envelopes} We begin by considering two basic geometric primitives to manipulate pseudoline families.

\begin{itemize}
    \item[(I)] \textbf{Point comparison.} 
    Given two $x,y\in X$ and a \emph{point} $s_i$, 
    decide if $f_x^s(s_i) \prec f_y^s(s_i)$. 
    \item[(II)] \textbf{Locating a crossing.} 
    Given two $x,y\in X$, 
    find the smallest $i$ such that $f_x^s(s_i) \succ f_y^s(s_i)$ or report that none exist. 
\end{itemize}
If these geometric primitives can be implemented in $\OO(1)$ time, we can compute the upper envelope with any algorithm, e.g. Graham's scan \cite{PreparataS85}. Graham's scan sorts the vertices of $X = \{x_1, \dots, x_{k}\}$ such that $f_{x_1}^s(s_1) \succ f_{x_2}^s(s_1) \succ \cdots \succ f_{x_{k}}^s(s_1)$, and then processes each pseudoline in this order from first to last, building the upper envelopes using a stack.
\begin{theorem}[Graham's Scan~\cite{PreparataS85}]\label{thm:grahams_scan}
There is an algorithm that takes as input $k$ pseudolines and computes the upper envelope of the pseudolines with $O(k\log k)$ many calls to point comparisons and locating crossings.
\end{theorem}

\paragraph*{Geometric primitives for distance profiles}
Here we show that point comparisons and locating a crossing can be performed in $\OO(1)$ time for any collection of $k$ distance profiles $\{f^s_x\}_{x\in X}$ for any of the $b$ choices of $s\in B$ (when we choose a different $s$ we rotate the labels on all the sites so $s$ appears first on the boundary ordering) on the boundary after $\OO(bk)$ preprocessing. This is notable because to specify a single distance profile requires $O(b)$ space, and there are $bk$ many pseudolines of the form $f^s_x$, so a naive implementation could require $O(b^2k)$ preprocessing. Conveniently, any fixed $x\in X$ and any $s_i, s_j\in B$, the distance profiles $f^{s_i}_x$ can be obtained from $f^{s_j}_x$ by shifting all coordinates by an additive constant $d_H(s_j,x)-d_H(s_i,x)$. 
Our following data structure is able to leverage this fact by exploiting the fact that this transformation preserves differences between adjacent points.
\begin{lemma}[Geometric Primitive Data Structure]\label{lem:geo_ops}
Let $H$ be a planar graph with a specified face $B$ and a set
$X\subseteq V(H)$ of $k$ target vertices, and let $b$ denote the
number of boundary vertices of $B$.
Suppose all distances from boundary vertices of $B$ to vertices
in $X$ are given.
After $\OO(bk)$ preprocessing, point comparisons and crossing-location queries for the family $\{f_x^s\}_{x\in X}$ for any fixed boundary vertex $s\in B$ can be answered in $\OO(1)$ time each.
\end{lemma}
\begin{proof}
The explicit distance matrix gives the $bk$ values of $f_x^s(s_i)$ for all $s_i\in B$ and $x\in X$.
It will be convenient for us to consider the discrete differences of adjacent values $D_x[s_i] = f_x^s(s_{(i-1)\pmod b}) - f_x^s(s_{i})$, where we define $s_0=s_b$. Observe that this does not depend on the choice of basepoint $s$, so there are at most $bk$ such values. 
After sorting these $bk$ values, we can translate the entries into rank space so all values are integers between $1$ and $bk$, without changing the answer to any comparison between them. 
\paragraph*{Point Comparison}
To handle point comparisons between $f_x^s(s_i)$ and $f_y^s(s_i)$, we can directly compute both values. If they are not equal, we are done.
Otherwise $f_x^s(s_i)=f_y^s(s_i)$,  we will lexicographically compare the following \emph{discrete difference strings}:
\begin{align*}
K_x^s(i) = (D_x[s_i], D_x[s_{i-1}],\dots,D_x[s_{2}])
\qquad \text{and} \qquad
K_y^s(i) = (D_y[s_i], D_y[s_{i-1}],\dots,D_y[s_{2}])
\end{align*}
Suppose $K_x^s(i) \neq K_y^s(i)$. Let $j$ be the largest value where $D_x[s_{j}] \neq D_y[s_j]$. Interchanging $x$ and $y$ if necessary, assume that $D_x[s_{j}] <D_y[s_j]$. As $f_x^s(s_i)=f_y^s(s_i)$, we have that 
\[f_x^s(s_j)=
f_x^s(s_i)+D_x[s_i]+D_x[s_{i-1}]+\cdots+D_x[s_{j+1}]= 
f_y^s(s_i)+D_y[s_i]+D_y[s_{i-1}]+\cdots+D_y[s_{j+1}]= 
f_y^s(s_j), \]
and that $f_x^s(s_{j-1}) < f_y^s(s_{j-1})$. Thus we can safely conclude that $f_x^s(s_{i}) \prec f_y^s(s_{i})$. 

Otherwise $K_x^s(i) = K_y^s(i)$ and all values of $f_x^s(s_\ell)= f_y^s(s_\ell)$ for all $1\le \ell \le i$. Then we must look ahead for a crossing:
\begin{align*}
\bar K_x^s(i) = (D_x[s_{i+1}], D_x[s_{i+2}],\dots, D_x[s_{b}])
\qquad \text{and} \qquad
\bar K_y^s(i) = (D_y[s_{i+1}], D_y[s_{i+2}],\dots, D_y[s_{b}])
\end{align*}
A similar argument allows us to decide if $f_x^s(s_{i}) \prec f_y^s(s_{i})$.

To handle these comparisons between strings, we consider the suffix array data structure \cite{ManberM93} that allows for lexicographical comparisons of arbitrary substrings (contiguous subsets) in $\OO(1)$ time.
Furthermore, we remark that for different base vertices $s\in B$, all strings of the form $K_x^s(i)$ (and $\bar K_x^s(i)$) are substrings of the (reversed) doubled discrete distance string:
\[\widetilde{D}_x = (D_x[s_b], \dots, 
D_x[s_1], 
D_x[s_b], 
\dots,
D_x[s_1]) 
\]
Thus we can build two single suffix array in $\OO(bk)$ time over the strings $\widetilde D_x$ and their reverse for all $x\in X$ to handle these queries for all boundary vertices $s\in B$ in $\OO(1)$ time.
\paragraph*{Locating a Crossing} There can be at most one crossing between any two pseudolines. If initially we have 
$f_x^s(s_1) \prec f_y^s(s_1)$, and for some later $i$ we have $f_x^s(s_i) \succ f_y^s(s_i)$, then $f_x^s(s_j) \succ f_y^s(s_j)$ for all $j>i$.
Thus we can binary search for the smallest $i$ with $O(\log b)$ calls to point comparison.
\end{proof}

Combining \Cref{lem:geo_ops} with \Cref{thm:grahams_scan} yields the following theorem. 
\begin{lemma}[Upper envelope data structure] \label{lem:upper-envelope}
Consider an $N$-vertex planar graph $H$ with a specified face $B$ and a set $X\subseteq V(H)$ of $k=|X|$ target vertices. 
Suppose all distances from boundary vertices $B$ to target vertices of $X$ are given.
After $\OO(bk)$ preprocessing,
given any boundary vertex $s\in B$ and a subset $X' \subseteq X$, 
the upper envelope of $\{f_x^{s}\}_{x\in X'}$ can be constructed in $\OO(|X'|)$ total time. The upper envelope can return $\max_{x\in X'} f_x^s(s_i)$ and a witness to the maximum in $O(\log |X'|)$ time.
\end{lemma}

\section{Farthest Vertex Data Structure}\label{sec:datastructure}

The main objective of this section is to prove the existence of the farthest vertex data structure of \Cref{thm:main_ds}. To do so, we will first describe a way to order the edges in \Cref{ssec:ccw-first}. We then show in \Cref{ssec:Voronoi_cells_and_sp} that it suffices to make a single data structure query with a contiguous interval in this ordering for each Voronoi edge of a Voronoi cell to find the farthest vertex.
\Cref{ssec:computing_voronoi_intervals} shows that we can compute such intervals efficiently.
\Cref{ssec:reductions} shows how the interval query can be handled by a reduction to computing pseudoline upper envelopes. Finally, we conclude by putting everything together to prove \Cref{thm:main_ds} in \Cref{ssec:wrapup}.

Recall that $H$ is an $N$-vertex planar graph with bidirected edges (possibly with different weights in each direction) and a collection of sites $B=(s_1, \dots, s_b)$ that lie in this order on the outer face of $H$. 
Let $T_i$ denote the shortest path tree rooted at $s_i$. 

\subsection{CCW-First Ordering}
\label{ssec:ccw-first}

We use the \emph{CCW-first ordering}
of~\cite{GawrychowskiKMS21} to assign distinct
indices to the vertices and directed edges of $H$.
For every directed edge $e=(u,v)$, add an artificial leaf
$x_e$ adjacent to $u$, placing its edge immediately before
$e$ in the counterclockwise order around $u$.
For a boundary root $s_i$, let $T_i^+$ consist of the
shortest path tree $T_i$ and all artificial leaf edges.

Perform a depth-first traversal of $T_i^+$, visiting the
children of each vertex in counterclockwise order, starting
immediately after the edge to its parent.
At $s_i$, start from a fixed imaginary incoming edge
embedded in the outer face.
Let $\operatorname{pre}_i(z)$ be the preorder index of a
vertex $z$ in this traversal, starting with
$\operatorname{pre}_i(s_i)=0$.
Define the \emph{CCW-first index} of a directed edge $e$ by $\rho_i(e)=\operatorname{pre}_i(x_e)$.
Every original vertex and every directed edge has a
distinct index, including both directions of tree and
non-tree edges.

We next show how to compute these indices for trees rooted at each boundary vertex $s_i\in B$ with only $\OO(N)$ total preprocessing.

\begin{lemma}[CCW-first indices]
\label{lem:ET-index}
After $\OO(N)$ preprocessing time, the CCW-first
index $\rho_i(e)$ and the preorder index
$\operatorname{pre}_i(v)$ can be reported in $\OO(1)$ time
for any boundary $s_i\in B$, directed edge $e$, and vertex $v \in V(H)$.
\end{lemma}

\begin{proof}
We use MSSP to maintain the shortest path tree as the source
moves around the outer face.
Following~\cite[Appendix~A.2]{CharalampopoulosGLMPWW23},
we use a fixed auxiliary root $r$ embedded in that face.
Add to $T_i$ an additional edge $(r,s_i)$, 
placed at the prescribed imaginary incoming edge of $s_i$.
MSSP performs $O(N)$ parent changes of the form
$ T\leftarrow T-\{(p,v)\}+\{(q,v)\}$,
where $q$ lies outside the subtree rooted at $v$.

We maintain the Euler tour of the augmented tree, traversing
each tree edge in both directions and following the
counterclockwise edge order at each vertex.
Store the Euler tour in a balanced binary tree supporting
splits, concatenations, and prefix sums.
Give value $1$ to edges moving away from $r$,
and value $0$ otherwise. These edges mark the first time a vertex $v\in V(H)$ or an artificial leaf $x_e$ is visited.
To compute $\operatorname{pre}_i(v)$, we take the prefix sum of the Euler tour before the traversal from its parent. The same query for the artificial leaf $x_e$ gives $\rho_i(e)$.

An MSSP parent change detaches the subtree rooted
at $v$ and inserts it at the position of $(q,v)$ in the
counterclockwise edge order at $q$.
Since the incoming edge at $v$ also changes, we cyclically
rotate the detached subtree tour to start immediately after
the new parent edge.
This requires only $O(1)$ splits and concatenations, together
with replacing the two occurrences of the old parent edge.
We also update the pointer to the first arrival at $v$.
Thus each change takes $O(\log N)$ time.

Finally, we make the binary tree storing the Euler tour persistent across the MSSP changes, and keep a version for every tree $T_i$ rooted at a boundary vertex $s_i\in B$. Since MSSP makes $O(N)$ tree changes, maintaining the Euler
tours takes $O(N\log N)$ time since each split and concatenation takes $O(\log N)$ time. Persistence costs an extra $O(\log N)$ factor in preprocessing time and query time. Thus the total preprocessing time and space are $\OO(N)$,
and each query takes $\OO(1)$ time.
\end{proof}

\paragraph*{Remark.}
The artificial leaves allow us to query $\rho_i(e)$ directly
even when $e$ does not belong to $T_i$.
For a fixed target vertex set $X$ of size $k$, we can also compute
$\operatorname{pre}_i(x)$ for all $s_i\in B$ and $x\in X$,
and sort $X$ in each ordering, in $\OO(bk)$ additional time.
Thus representing the orderings and extracting their
restrictions to $X$ takes $\OO(N+bk)$ total preprocessing
time and space.

\subsection{Voronoi Cells and Shortest Path Trees}
\label{ssec:Voronoi_cells_and_sp}
We will discuss the interaction between the Voronoi cell $\Vor(s)$ for some $s=s_i\in B$ in the Voronoi diagram $\VD(H, B, \omega)$ with respect to arbitrary admissible weights $\omega$, the shortest path tree $T_i$, and a CCW-first index of the edges. 
Let $E_1, \dots, E_d$ be the edges of the reduced dual Voronoi diagram of $\Vor(s)$ in counterclockwise order around the Voronoi cell. Each reduced Voronoi edge consists of a set of dual edges that separate $\Vor(s)$ from another Voronoi cell $\Vor(s_{j_\ell})$. 
Let $e_\ell=(u_\ell,v_\ell)$ be the directed primal edge whose dual is the first edge of $E_\ell$ encountered in a counterclockwise walk around the boundary of $\Vor(s)$, oriented so that $u_\ell\in\Vor(s)$ and $v_\ell\in\Vor(s_{j_\ell})$.
Let $J_\ell$ consist of the original vertices $x$ whose positions $\pre_i(x)$ lie in the cyclic interval $[\rho_i(e_\ell),\rho_i(e_{\ell+1}))$, where the interval for the last reduced edge $E_d$ ends at $\rho_i(e_1)$.
If $d=1$, this interval comprises one complete traversal.
The following lemma relates these intervals to the structure of $\Vor(s)$ in $T_i$ and shows that, within each $J_\ell$, membership in $\Vor(s)$ is determined by a single comparison with the neighboring site $s_{j_\ell}$.

\begin{lemma}\label{lem:CCW-first-voronoi}
Fix a site $s=s_i$, and let $T$ be a shortest path tree rooted at $s$. Then the following is true.
\begin{enumerate}
    \item[(a)]  \textbf{Connected Voronoi subtree.} \cite[Lemma 2.1]{GawrychowskiKMS21}
    The vertices of $\Vor(s)$ form a connected induced subtree of $T$.
    \item[(b)] \textbf{Voronoi edges give CCW-first intervals.}
    Let $C_e$ be a connected subtree of $T-V(\Vor(s))$, with parent edge $e$ directed from $\Vor(s)$ into $C_e$, whose dual belongs to $E_\ell$.
    The entire CCW-first traversal into $C_e$ is a cyclic interval contained in the cyclic interval $[\rho_i(e_\ell),\rho_i(e_{\ell+1}))$ with vertices $J_\ell$.
    In particular, $V(C_e)\subseteq J_\ell$.
    \item[(c)] \textbf{One comparison per interval.}
    For every $x\in J_\ell$, $x\in\Vor(s)$ if and only if $f_x^s(s_{j_\ell})\geq\omega(s)-\omega(s_{j_\ell})$.
\end{enumerate}
\end{lemma}
\begin{proof}
We begin by showing that normalized distance profiles are monotone non-increasing as we walk down the shortest path tree. Consider a vertex $v$ with parent $p$ in $T$. 
As the edge $e = pv$ lies on the shortest path from $s$ to $v$, we have that $d_H(s,v) = d_H(s, p) + w(e)$. For any other site $t$, triangle inequality gives $d_H(t,v) \le d_H(t, p) + w(e)$. Subtracting this yields:
\begin{equation} \label{eq:monotone}
   f_v^s(t) = d_H(t,v) - d_H(s,v) \le d_H(t,p) -d_H(s, p) = f_p^s(t). 
\end{equation} 

\paragraph*{Proof of (a) connected Voronoi subtrees.}
A vertex $x\in V(H)$ is in $\Vor(s)$ if $x$ is closer to $s$ than any other site. This means that $f_x^s(t) \ge \omega(s)-\omega(t)$ for every other $t\in B$. Thus if $x$ has parent $p$ in $T$, then for every $t\in B$ we must have $f_p^s(t) \ge f_x^s(t) \ge \omega(s)-\omega(t)$.
By induction, the entire shortest path from $s$ to $x$ is in $\Vor(s)$. Thus we conclude that $\Vor(s)$ is a connected subtree.

\paragraph*{Proof of (b) Voronoi edges give CCW-first intervals.}

Let $\Gamma$ be the boundary walk of $\Vor(s)$ in the dual.
By part~(a), the paths in $T$ from $s$ to the tails of the edges leaving $\Vor(s)$ lie entirely in $\Vor(s)$.
These paths are non-crossing, and the CCW-first search explores their branches in counterclockwise order.
Thus, contracting the subtree of $T$ induced by $\Vor(s)$ to $s$ preserves the cyclic order in which the search encounters the outward boundary edges.
After contraction, these edges occur counterclockwise around $s$ in the order of their duals along $\Gamma$, so the cyclic order of the reduced Voronoi edges agrees with the CCW-first order.

Each reduced Voronoi edge $E_\ell$ is a contiguous subpath of $\Gamma$. 
Thus the outward directed primal edges whose duals belong to $E_\ell$ occur consecutively among the outward boundary edges encountered by the CCW-first search.
Say $e^*\in E_\ell$ is an edge dual to the primal edge $e$ of $T$ and $C_e$ is the connected component of $T-\Vor(s)$.
Once the tour enters $C_e$ through the edge $e$, it visits the entire rooted subtree $C_e$ before returning to $\Vor(s)$. 
Every interval endpoint is the index of an artificial leaf attached to a vertex of $\Vor(s)$, so no endpoint is encountered during this excursion.
Since $\rho_i(e)$ lies in $[\rho_i(e_\ell),\rho_i(e_{\ell+1}))$ and is visited immediately before entering $C_e$, the entire CCW-first traversal of $C_e$ lies in this cyclic interval. See \Cref{fig:voronoi-euler-intervals} for an illustration of this.

\begin{figure}[h]
    \centering
    \includegraphics[width=0.8\linewidth]
        {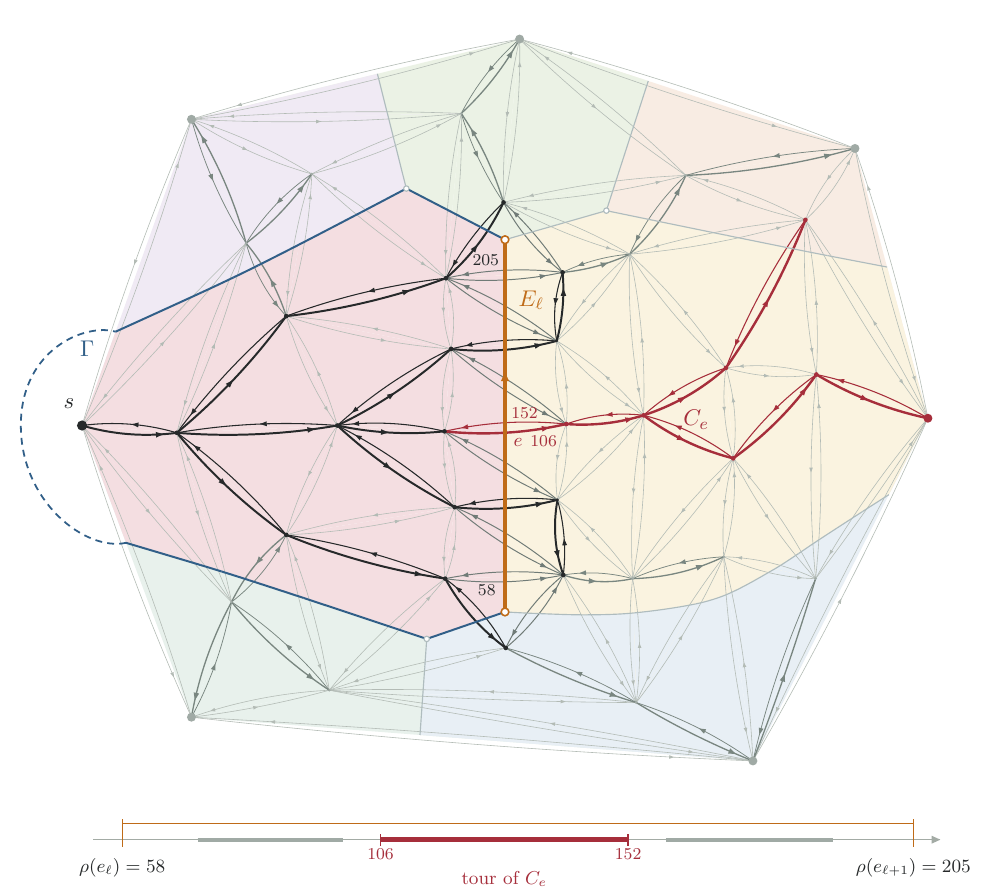}
    \caption{Voronoi edges give CCW-first intervals.
    The pink cell $\Vor(s)$ has dual boundary $\Gamma$, with the reduced Voronoi edge $E_\ell$ highlighted in orange.
    Paired arcs show the two traversal directions of each primal edge.
    The labels $\rho_i(a)$ are the CCW-first preorder indices of the artificial leaves associated with the directed arcs $a$.
    Numbering starts at $s$ with index $0$, using the leftward ray in the outer face as reference.
    Every original vertex and every artificial leaf contributes one index, including leaves for reverse and non-tree arcs.
    The red tree edge $e$ enters the component $C_e$ of $T-V(\Vor(s))$, where $T$ is the shortest path tree rooted at $s$.
    On the bottom line, the orange bracket marks $[\rho_i(e_\ell),\rho_i(e_{\ell+1}))=[58,205)$.
    The red segment marks the complete excursion into $C_e$, from $\rho_i(e)=106$ to the reverse arc at index $152$; gray segments mark the other excursions through $E_\ell$.
    The augmented subtree of $C_e$ contains seven original vertices and $39$ artificial leaves, accounting for the difference $152-106=7+39$.
    Thus $V(C_e)\subseteq J_\ell$.
    The dashed arc closes $\Gamma$ through the outer face.}
    \label{fig:voronoi-euler-intervals}
\end{figure}

\paragraph*{Proof of (c) one comparison per interval.}
Write $t=s_{j_\ell}$, and consider a vertex $x\in J_\ell$. Suppose $x\in \Vor(s)$, then clearly $f_x^s(t) \ge \omega(s) - \omega(t)$. On the other hand if $x\not\in \Vor(s)$, then by (a) $x$ belongs to a rooted subtree $C_e$ of $T-V(\Vor(s))$ with parent edge $e$. Let $v$ be the endpoint of $e$ that is the root of $C_e$. 
By part~(b), the dual of $e$ belongs to $E_\ell$.
Since $e$ separates $\Vor(s)$ from $\Vor(t)$, we must have $v\in \Vor(t)$. This means that $f_v^s(t) < \omega(s)-\omega(t)$. By \Cref{eq:monotone}, it follows that:
\[ f_x^s(t) \le f_v^s(t) < \omega(s)-\omega(t).\]
This means $x\not\in \Vor(s)$ fails the comparison, proving (c).
\end{proof}

\subsection{Computing Voronoi Edge Intervals}
\label{ssec:computing_voronoi_intervals}

We use a theorem of Mozes and Prigan~\cite{MozesP2026} that efficiently computes a Voronoi diagram of a set $B$ of sites lying on one face. We can assume all other faces have exactly three sides without changing the underlying graph metric by adding long edges (see simplifying assumptions of \Cref{sec:preliminaries}). The key idea is to find \emph{trichromatic triangles} which are faces with three vertices belonging to different Voronoi cells. These are the dual vertices of the Voronoi diagram of degree three. The theorem is as follows.
\begin{theorem}[Lemma 3.1 of~\cite{MozesP2026}] \label{thm:tritriangle}
Let $H$ be an $N$-vertex directed planar graph with real weights and no negative weight cycles, and $B$ be a collection of $b$ sites on one face. Assume that all other faces of $H$ have degree $3$. After $\widetilde O(N)$-time preprocessing, for any admissible weight function $\omega$, an implicit representation of $\VD(H, B, \omega)$ along with all trichromatic triangles can be computed in $\OO(b)$ time.
\end{theorem}
To recover the intervals according to the CCW-first indices of the shortest path trees, we can use \Cref{lem:ET-index} to get the index $\rho(e)$ for trichromatic edges.

Together with the compressed Voronoi diagram and \Cref{lem:ET-index}, we can compute the indices of the interval corresponding to each Voronoi edge. 
\begin{corollary}\label{lem:VorET-interval}
After $\OO(N)$ preprocessing, for every admissible weight function $\omega$ on a set of $b$ boundary sites $B'\subseteq B$, 
after computing the implicit Voronoi diagram in $\OO(b)$ time and trichromatic triangles with \Cref{thm:tritriangle}, we can recover the induced CCW interval of a Voronoi edge in $\OO(1)$ time.
\end{corollary}

\subsection{Data Structures via Range Searching} \label{ssec:reductions}
In this section we describe standard range searching data structures to show that finding the farthest vertex can be performed using upper envelopes of distance profile pseudolines.

\Cref{lem:CCW-first-voronoi}(c) says that we can decompose the Euler tour into intervals where we need to only compare distance profiles with a single competitor boundary site to determine if vertices are in $\Vor(s)$.
This means that for interval $J_\ell$, it suffices to query $|\Vor(s)|$ many intervals for the vertex $x$ that maximizes $d_H(s,x)$ while satisfying $f_x^s(s_{j_\ell}) \ge \omega(s) - \omega(s_{j_\ell})$.
This can be encapsulated in the following problem.

\begin{problem}[Interval Threshold Maximization Problem] \label{prob:max}
Consider a sequence of $k$ pseudolines $f_1, f_2, \dots, f_k$ that take on values on $b$ unique coordinates where each pseudoline has an associated distance $d_1, d_2, \dots, d_k$. 
Construct a data structure to support queries of an interval $I$, a pseudoline coordinate $t$, and a threshold $\tau$ to output the index $i$ that maximizes $d_i$ subject to $i\in I$ such that $f_i(t) \ge \tau$.
\end{problem}

This problem can be solved using standard multilevel orthogonal range searching technique where we store the upper envelope data structures. 
We give here the full proof of this via reduction to a problem that can be solved using upper envelopes (in our case with \Cref{lem:upper-envelope}). 
For this reduction, it helps to define the following intermediate problem.
\begin{problem}[Interval Threshold Decision Problem] \label{prob:decision}
Consider a sequence of $k$ pseudolines $f_1, f_2, \dots, f_k$ that takes on values on $b$ unique coordinates.
Construct a data structure to support queries of an interval $I$, a pseudoline coordinate $t$, and a threshold $\tau$ to an index $i\in I$ where $f_i(t) \ge \tau$.
\end{problem}
We also define the following problem that can be solved using upper envelopes.
\begin{problem}[Global Threshold Decision Problem] \label{prob:decision2}
Consider a sequence of $k$ pseudolines $f_1, f_2, \dots, f_k$ that takes on values on $b$ unique coordinates.
Construct a data structure to support queries for a pseudoline coordinate $t$ and a threshold $\tau$ to output any index $i$ where $f_i(t) \ge \tau$.
\end{problem}

The reductions are standard. We present a full proof here for completeness. 
\begin{lemma}[Data structure reductions]\label{lem:reductions}
Suppose a data structure for \Cref{prob:decision2} can be constructed in $T_3(k)$ time and could handle queries in $Q_3(k)$ time. Assuming that $T_3(k)=\Omega(k)$, a data structure for \Cref{prob:max} can be constructed in $\OO(T_3(k))$ time with $\OO(Q_3(k))$ query time.
\end{lemma}

\newcommand{\lo}{\textsc{low}}
\newcommand{\hi}{\textsc{high}}
\begin{proof}
As an intermediate step, we first reduce \Cref{prob:max} to \Cref{prob:decision}.
\paragraph*{Reducing from \Cref{prob:max} to \Cref{prob:decision}}
We proceed by standard divide and conquer. Sort all values of $d_i$, and split the pseudolines into two sets $F_\hi$ and $F_\lo$ for which all values in $F_\hi$ are greater than those in $F_\lo$. Build the data structure for \Cref{prob:decision} on $F_\hi$, then recursively build a data structure for \Cref{prob:max} on both $F_\lo$ and $F_\hi$. 
Queries can be handled by binary search, use the data structure for $F_\hi$ to determine if any pseudoline with $d_i$ value larger than $m$ satisfies the interval and threshold condition. If there is, then recurse on $F_\hi$, otherwise recurse on $F_\lo$. In the recursion, binary search for new endpoints of the query interval, since $F_\lo$ and $F_\hi$ only contain subsets of the original indices.

Suppose we had a data structure for \Cref{prob:decision} with $T_2(k)$ preprocessing time with $Q_2(k)$ query time. The construction time for the data structure for \Cref{prob:max} has the recurrence $T_1(k) = T_2(k/2)+2T_1(k/2) + O(k\log k)$. This solves to $T_1(k) = \OO(T_2(k))$ as any data structure for $T_2(k)$ takes at least $\Omega(k)$ time. 
The query time has the recurrence $Q_1(k) = Q_2(k/2)+Q_1(k)+O(\log k)$, which solves to $Q_1(k) = \OO(Q_2(k))$.

\paragraph*{Reducing from \Cref{prob:decision} to \Cref{prob:decision2}}
This follows by standard segment trees \cite{deBergCKO08}, that decompose an interval into $O(\log k)$ dyadic intervals. 
Suppose we had a data structure for \Cref{prob:decision2} with $T_3(k)$ preprocessing time with $Q_3(k) = \OO(1)$ query time. Then the preprocessing time satisfies $T_2(k) = 2T_2(k/2)+T_3(k)$ which solves to $T_2(k) = \OO(T_3(k))$ by building a solution to \Cref{prob:decision2}, and ignoring the global preprocessing time. To perform a query, it takes $Q_2(k) = O(\log k)\cdot Q_3(k)$ time.
Thus we have shown that $T_1(k) = \OO(T_2(k)) = \OO(T_3(k))$ and that $Q_1(k) = \OO(Q_2(k)) = \OO(Q_3(k))$. so we conclude the lemma.
\end{proof}

\subsection{Proving \Cref{thm:main_ds}}\label{ssec:wrapup}

Here we prove our main data structure result of \Cref{thm:main_ds} that we restate below for convenience.
\mainds*
\begin{proof}
We begin by describing the data structure.

\paragraph*{Data structure}
We store the data structure of \Cref{lem:VorET-interval}
for recovering Voronoi edge intervals and a single MSSP
data structure from \Cref{thm:mssp} shared by all boundary
vertices.
For each $s_i\in B$, we also store the data structure for
\Cref{prob:max} on the pseudolines $\{f_x^{s_i}\}_{x\in X}$ in increasing order of $\pre_i(x)$, associating
each pseudoline $f_x^{s_i}$ with the distance
$d_H(s_i,x)$.

\paragraph*{Preprocessing time}
Constructing the data structures for computing Voronoi edge intervals and MSSP takes $\OO(N)$ preprocessing time.
Using distances obtained by the MSSP, \Cref{lem:upper-envelope} allows for constructing an upper envelope to solve the decision problem of  \Cref{prob:decision2} on any $k'$ pseudolines with the same boundary vertex $s_i\in B$ in $\OO(k')$ time after $\OO(bk)$ preprocessing. 
The reduction of \Cref{lem:reductions} therefore gives
$\OO(k)$ construction time for the data structure for
\Cref{prob:max} at each boundary vertex, or $\OO(bk)$
over all boundary vertices.
The total preprocessing time is thus $\OO(N+bk)$.

\paragraph*{Handling a query}
Given an admissible weight function $\omega$, we use
\Cref{lem:VorET-interval} to construct an implicit
representation of its Voronoi diagram in $\OO(b)$ time.
For each Voronoi edge on the boundary of a cell $\Vor(s_i)$,
we recover the corresponding interval in the CCW-first
ordering rooted at $s_i$ in $\OO(1)$ time.
By \Cref{lem:CCW-first-voronoi}(c), processing this interval
requires one query (or two for circular intervals) 
to the data structure for \Cref{prob:max}
associated with $s_i$, also taking $\OO(1)$ time.
Thus, after constructing the diagram, we can process all
boundary edges of $\Vor(s_i)$ in $\OO(|\pVor(s_i)|)$ time.
\end{proof}

\section{Conclusion}
We have reached a milestone for planar diameter and all vertex eccentricities by showing that these quantities can be computed in $\OO(n^{3/2})$ time. 
We conclude with some open problems.
\begin{enumerate}
    \item \textbf{Tighter bounds for exact planar diameter.} 
    We reiterate the question asked by Abboud, Mozes, and Weimann~\cite{AbboudMW23}. Does planar diameter admit an $n^{1+o(1)}$-time algorithm or can a superlinear lower bound be proved? Are other improvements beyond $\OO(n^{3/2})$ for planar diameter possible?
    \item \textbf{Wiener index.} The Wiener index computes the sum of shortest path distances between all pairs of vertices. Upper envelopes are only good for maximum queries, so our farthest vertex data structure of \Cref{thm:main_ds} does not naturally support computation of Wiener index. Typically, supporting the Wiener index requires computation of sums and counts of vertices in a Voronoi cell. Since queries can be processed offline, it may be possible to use point-pseudoline range searching data structures to handle $q$ queries on $k$ pseudolines in $\OO(q^{2/3}k^{2/3})$ time. Unfortunately this does not directly lead to any improvement on the $\OO(n^{5/3})$ time algorithm in \cite{GawrychowskiKMS21}.
    \item \textbf{Dynamic exact diameter and distance oracles.}
    Can the structural observations relating to pseudolines made in this paper help for dynamic exact distance problems?
\end{enumerate}

\subsection*{AI Methodology}
The results presented in this paper were developed through multiple interactions with GPT-6 Astra. The idea to use range searching and pseudolines was suggested by the first author. The manuscript itself was written by the authors with editing assistance by the LLM.

\bibliography{ref}
\bibliographystyle{alpha}


\end{document}